# Poster: A general practitioner or a specialist for your infected smartphone?

Jelena Milosevic
Advanced Learning and Research Institute (ALaRI)
Faculty of Informatics
Università della Svizzera italiana
Lugano, Switzerland
Email: jelena.milosevic@usi.ch

Alberto Ferrante
ALaRI
Faculty of Informatics
Università della Svizzera italiana
Lugano, Switzerland
Email: alberto.ferrante@usi.ch

Miroslaw Malek
ALaRI
Faculty of Informatics
Università della Svizzera italiana
Lugano, Switzerland
Email: miroslaw.malek@usi.ch

*Abstract*—With explosive growth in the number of mobile devices, the mobile malware is rapidly spreading as well, and the number of encountered malware families is increasing. Existing solutions, which are mainly based on one malware detector running on the phone or in the cloud, are no longer effective. Main problem lies in the fact that it might be impossible to create a unique mobile malware detector that would be able to detect different malware families with high accuracy, being at the same time lightweight enough not to drain battery quickly and fast enough to give results of detection promptly.

The proposed approach to mobile malware detection is analogous to general practitioner versus specialist approach to dealing with a medical problem. Similarly to a general practitioner that, based on indicative symptoms identifies potential illnesses and sends the patient to an appropriate specialist, our detection system distinguishes among symptoms representing different malware families and, once the symptoms are detected, it triggers specific analyses. A system monitoring application operates in the same way as a general practitioner. It is able to distinguish between different symptoms and trigger appropriate detection mechanisms. As an analogy to different specialists, an ensemble of detectors, each of which specifically trained for a particular malware family, is used. The main challenge of the approach is to define representative symptoms of different malware families and train detectors accordingly to them. The main goal of the poster is to foster discussion on the most representative symptoms of different malware families and to discuss initial results in this area obtained by using Malware Genome project dataset.

## I. Introduction

As smartphone has become the most popular device on the planet, its security is fundamental, having direct impact on productivity and well-being of millions of individuals, companies, and governments. However, the implications on security of rapid deployment coupled with always-on connectivity are still insufficiently understood. That is why threat alerts constantly grew year by year, and according to McAfee Labs [1] the collection of mobile malware continued its steady climb as it broke 6 million samples in the forth quarter of 2014, up 14% over the third quarter of the same year.

Any security mechanism targeted toward mobile systems must take their limitations into consideration as they may significantly limit the ability to run complex malware detection systems on the device. Furthermore, if a user notices that the malware detection system drains battery quickly or slows down the operating system, chances are high that he/she will turn it off and leave the device unprotected.

The methodology that we propose is taking the mentioned limitations into account by using a lightweight symptom monitoring infrastructure that is able to detect suspicious symptoms, and triggers a dedicated, more precise but possibly more complex detector that is previously trained to identify the malware family with higher confidence. Although the goal of the proposed methodology is to develop a comprehensive malware detection system following the mentioned approach, the main focus of the poster is to discuss its first part: the symptom monitoring infrastructure. We would like to foster discussion on different features that can be collected and observed on the smartphone, their connection to different malware families and relevance to the detection accuracy.

## II. State of The Art

In order to extract the most indicative symptoms, feature selection and extraction has to be done. Broad survey covering the state of the art in feature selection can be found in [2].

In [3] the authors propose to identify malware with sets of permissions. The authors concluded that although the number of permissions alone is not sufficient to identify malware, they could be used as part of a set of classification features, provided that all permissions common to the malware set are infrequent among non-malicious applications. In [4], as a feature for detecting likelihood of malware infection, type of applications running on a device is used. While observing just this feature is not enough to give precise answer about device being attacked, we believe that using it as one of the indicators could give good results. As outlined in [5], several solutions rely on observation of battery power. However, to what extent malware can be detected on smartphones monitoring just the battery power remains an open research question, having in mind continuously changing user behavior.

## III. Proposed Methodology and Initial Results

The system monitoring application, running on the phone, collects information related to its condition, and upon suspicious symptoms detection triggers more specific detection. The outline of the approach is presented in Figure 1. More detailed description of the complete methodology can be found in [6]. In order to identify the most indicative symptoms, our idea is to combine features from state-of-the-art (sets of permissions, likelihood of malware infection, and battery power) together with memory consumption information, processor usage and network behaviour. By applying statistical methods, we will define the most representative ones with respect to different malware families. During the training phase, malware samples from different families are executed, relevant data collected and afterwards analysed. At runtime, the symptoms monitoring infrastructure collects information about the phone and, based on their values and previously defined symptoms, triggers the appropriate, more specific detection.

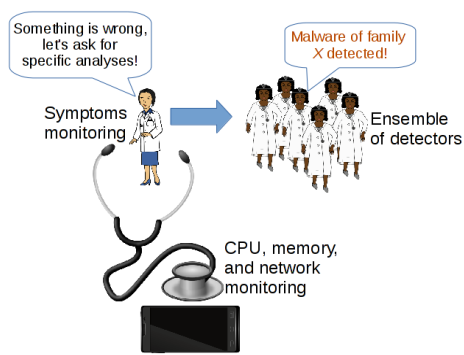

Fig. 1. A representation of the proposed approach

TABLE I. THE MOST RELEVANT FEATURES PER MALWARE FAMILY

| Gold Dream | Geinimi | Base Bridge |
|---|---|---|
| Page Major Faults | Page Major Faults | Page Major Faults |
| Cursor Pss | Cursor Pss | Cursor Private Dirty |
| Dalvik Heap Free | Dalvik Heap Free | Cursor Pss |
| .so mmap Shared Dirty | .so mmap Private Dirty | Dalvik Heap Alloc |
| Cursor Private Dirty | Cursor Private Dirty | Ashmed Pss |
| **Fake Player** | **jSMShider** | **Pjapps** |
| Page Minor Faults | Page Major Faults | Page Major Faults |
| Dalvik Heap Free | Ashmed Pss | Cursor Pss |
| Dalvik Heap Alloc | Dalvik Heap Free | Dalvik Heap Free |
| .so mmap Pss | Cursor Pss | Page Minor Faults |
| Dalvik Private Dirty | Ashmed Shared Dirty | Cursor Private Dirty |

Initial results have been obtained by analysing the following malware families, taken from the Malware Genome project dataset [7]: **Gold Dream** that spies on SMS messages received by users so as on incoming/outgoing phone calls and then uploads them to a remote server without user's awareness; **Geinimi** Trojan that compromises personal data on a user's phone and sends it to the remote servers (it sends location coordinates, device identifiers, so as a list of the installed apps on the phone); **Base Bridge** upon activation it communicates with a control server, dials calls or sends SMS messages (it also blocks messages from the mobile carrier to prevent users from getting consumption fee updates on time); **Fake Player** that pretends to be a movie player, but instead sends SMS messages to present numbers; **jSMShider** Trojan that affects devices where the owner has downloaded a custom ROM or rooted phone; and **Pjapps** Trojan that opens a back door on the compromised device and retrieves command from a remote command and control server.

Malware samples (318 apps) belonging to the six mentioned families have been executed in the Android SDK, virtual device running Android 4.0 (API 14). For the purpose of this poster, only the importance of features related to CPU and memory are reported, since, to the best of our knowledge, such analysis have not been done before. CPU activity is described in more detail by the following features: total CPU usage, user CPU usage, kernel CPU usage, page minor and major faults. To investigate memory usage, all features related to apps memory allocation obtained by using "adb shell dumpsys meminfo <package_name>" command have been collected.

Features have been analysed with Principal Component Analysis (PCA) method using Ranker search in the Weka data mining tool [8]. Variance is set to 95%. Attribute noise is filtered by transforming to the PCA space, eliminating some of the worst eigenvectors, and then transforming back to the original space. Relevance of the features is given with respect to the transformed features space, as opposed to regular ranking methods that work with the original space; obtained results are reported in Table I. Initial results show that different families have different representative features. However, some features have high importance in every transformed space: Page Major Faults, Cursor Proportional set size, Dalvik Heap Free, related to CPU consumption, sharing pages across processes, Dalvik allocations, respectively. As a next step towards the identification of the most indicative symptoms, we plan to further analyse existing families by clustering them, based on aforementioned features.

IV. CONCLUSION

The poster presents a new approach to mobile malware detection, by using general practitioner versus specialist approach from medical field. Symptoms monitoring application (an equivalent of a general practitioner) is running on the phone checking for the symptoms of malware (illnesses), and upon a detection of a suspicious symptoms triggers appropriate, more detailed, analysis (an equivalent of an appropriate specialist). The main focus is on the selection of the most indicative symptoms for different malware families. The initial results towards the solution are presented and discussed. Analysed malware families are obtained from Malware Genome Dataset.